\documentclass[]{spie}  

\usepackage{amsmath,amsfonts,amssymb}
\usepackage{graphicx}
\usepackage{url}
\usepackage[colorlinks=true, allcolors=blue]{hyperref}

\title{Towards independent event horizon imaging of the supermassive black holes in M87 and the Milky Way}

\author[a]{Nithyanandan Thyagarajan}
\author[a]{Samuel Lai}
\author[a]{O. Ivy Wong}
\author[b]{Foivos Diakogiannis}
\affil[a]{Space \& Astronomy, Commonwealth Scientific and Industrial Research Organisation (CSIRO), P. O. Box 1130, Bentley, WA 6102, Australia}
\affil[b]{Data 61, Commonwealth Scientific and Industrial Research Organisation (CSIRO), Kensington, WA 6151, Australia}

\authorinfo{Further author information: (Send correspondence to N.T.)\\N.T.: E-mail: Nithyanandan.Thyagarajan@csiro.au, Telephone: +61 8 6436 8626}

\begin{document} 
\maketitle

\begin{abstract}
The Event Horizon Telescope (EHT) Collaboration's images of the supermassive black holes in M87 and the Milky Way have provided the first event-horizon-scale views of these objects, opening new avenues for studies of gravitation, accretion physics, and black hole astrophysics. Achieving these results, however, requires imaging under some of the most challenging conditions in radio astronomy, including low signal-to-noise ratios, severe calibration uncertainties, and sparse aperture coverage.

With the aim of presenting independent analyses of the public EHT datasets for M87* and Sgr~A*, we adopt an approach that is independent in observables, and reconstruction methodology. Our framework is based on closure invariants, a class of interferometric observables that are intrinsically immune to station-based calibration errors and therefore provide robust constraints on source structure. We combine these observables with Generative Deep learning Image Reconstruction with Closure Terms (GenDIReCT), a diffusion-based image reconstruction framework that operates in the latent space of images conditioned on closure invariants.

We present independent reconstructions obtained using GenDIReCT on synthetic challenge data sets as well as real EHT data on 3C279 and Centaurus~A, and compare them with previously reported results. This work demonstrates the potential of closure-invariant-driven generative imaging as a calibration-resilient framework for Very Long Baseline Interferometry (VLBI) and provides an independent and complementary avenue for interpreting horizon-scale black hole observations.

\end{abstract}


\section{INTRODUCTION}
\label{sec:intro}  

Very Long Baseline Interferometry (VLBI) \cite{TMS} is a radio interferometric technique that achieves the very high angular resolution in astronomy by combining signals from widely separated telescopes. By synthesising an aperture comparable in size to the Earth, VLBI enables imaging of astrophysical phenomena on event-horizon scales that are inaccessible to conventional instruments.

The Event Horizon Telescope (EHT) \cite{Doeleman_2009} is a global VLBI array designed to image the immediate environments of supermassive black holes. By linking telescopes across multiple continents, the EHT attains the angular resolution required to resolve structures near the event horizons of the black holes in M87 and the Milky Way. The resulting images \cite{EHT_2019_IShadow,EHT_2022_SgrAImaging} have become iconic demonstrations of horizon-scale imaging and are widely regarded as supporting the predictions of general relativity in the strong-field regime.

Despite these successes, VLBI imaging remains an inherently challenging inverse problem. The observations are characterised by extremely sparse aperture coverage, low signal-to-noise ratios, and significant calibration uncertainties arising from instrumental and propagation effects at geographically dispersed stations \cite{TMS}. Consequently, image reconstruction often depends on a range of algorithmic choices and hyperparameters, including initial models, regularisation strategies, deconvolution windows, and assumed fields of view. Differences in these choices have led to alternative image reconstructions and, in some cases, interpretations that differ from the published EHT results \cite{Miyoshi_2022_M87,Miyoshi_2024}.

Given the scientific significance of horizon-scale black hole imaging, independent analyses based on fundamentally different assumptions and methodologies are essential. In this work, we adopt an approach based on closure invariants \cite{Jennison_1958,Twiss_1960,Broderick_2022,Thyagarajan_2022_CPhase,Thyagarajan_2022_CI,Samuel_2022}, a family of interferometric observables that are intrinsically immune to station-based calibration errors. By construction, closure invariants isolate source-structure information while mitigating one of the dominant sources of systematic uncertainty from calibration in VLBI data analysis.

To reconstruct images from these observables, we employ Generative Deep learning Image Reconstruction with Closure Terms (\texttt{GenDIReCT}), a generative diffusion framework conditioned directly on closure invariants. Operating in the latent space of images, \texttt{GenDIReCT} leverages learned image priors while reducing dependence on many of the arbitrary imaging hyperparameters that commonly influence traditional reconstruction methods. This combination of calibration-independent observables and generative modelling provides a complementary pathway for interpreting horizon-scale VLBI observations.

The paper is organised as follows. Section~\ref{sec:closure-invariants} introduces the concept of closure invariants. Section~\ref{sec:gendirect} describes the deep learning image reconstruction framework using diffusion conditioned on closure invariants. Current status and results are provided in section~\ref{sec:results}. Future outlook and summary are presented in section~\ref{sec:summary}.

\section{Closure Invariants}\label{sec:closure-invariants}

VLBI operates under some of the most challenging observational conditions in astronomy. The large separations between heterogeneous telescope elements lead to sparse aperture coverage and significant calibration uncertainties, with each station subject to distinct instrumental characteristics and environmental effects. To mitigate these challenges, VLBI has long relied on closure quantities—most notably closure phases \cite{Jennison_1958,Thyagarajan_2022_CPhase} and closure amplitudes \cite{Twiss_1960}. These observables are formed from specific combinations of visibilities that are intrinsically immune to station-based gain corruptions, enabling robust inference of source structure even when accurate calibration is difficult.

Closure quantities have been central to many significant advances in high-resolution radio astronomy. Early applications led to the discovery of the double-lobed morphologies of Cygnus A \cite{Jennison1957,Jennison+Latham1959} and Centaurus A \cite{Twiss_1960}, while later studies used them to establish the core--jet structure of quasar 3C~147 \cite{Wilkinson_1977} and provide the first direct evidence for superluminal motion in the jet of 3C~273 \cite{Pearson_1981}. More recently, closure quantities played a critical role in the Event Horizon Telescope (EHT) imaging campaigns of the supermassive black holes in M87* \cite{EHT_2019_IShadow} and Sgr A* \cite{EHT_2022_SgrAImaging}, where they were incorporated alongside calibrated visibilities in horizon-scale image reconstruction.

\section{Generative Deep learning Image Reconstruction from Closure Terms (\texttt{GenDIReCT})}\label{sec:gendirect}

Building on this long history, recent theoretical and algorithmic developments have demonstrated that closure quantities alone contain sufficient information to enable image reconstruction without reliance on station gain calibration. Examples include regularised maximum likelihood methods such as \texttt{eht-imaging} \cite{Akiyama_2017_SMILI,Chael_2018_ehtim} and parametric imaging \cite{Thyagarajan_2024_Lucas}. This has opened a new avenue for calibration-independent imaging, allowing source structure to be inferred directly from observables that are fundamentally immune to one of the dominant sources of systematic uncertainty in VLBI.

\begin{figure} [ht]
   \begin{center}
   \begin{tabular}{c} 
   \includegraphics[height=7cm]{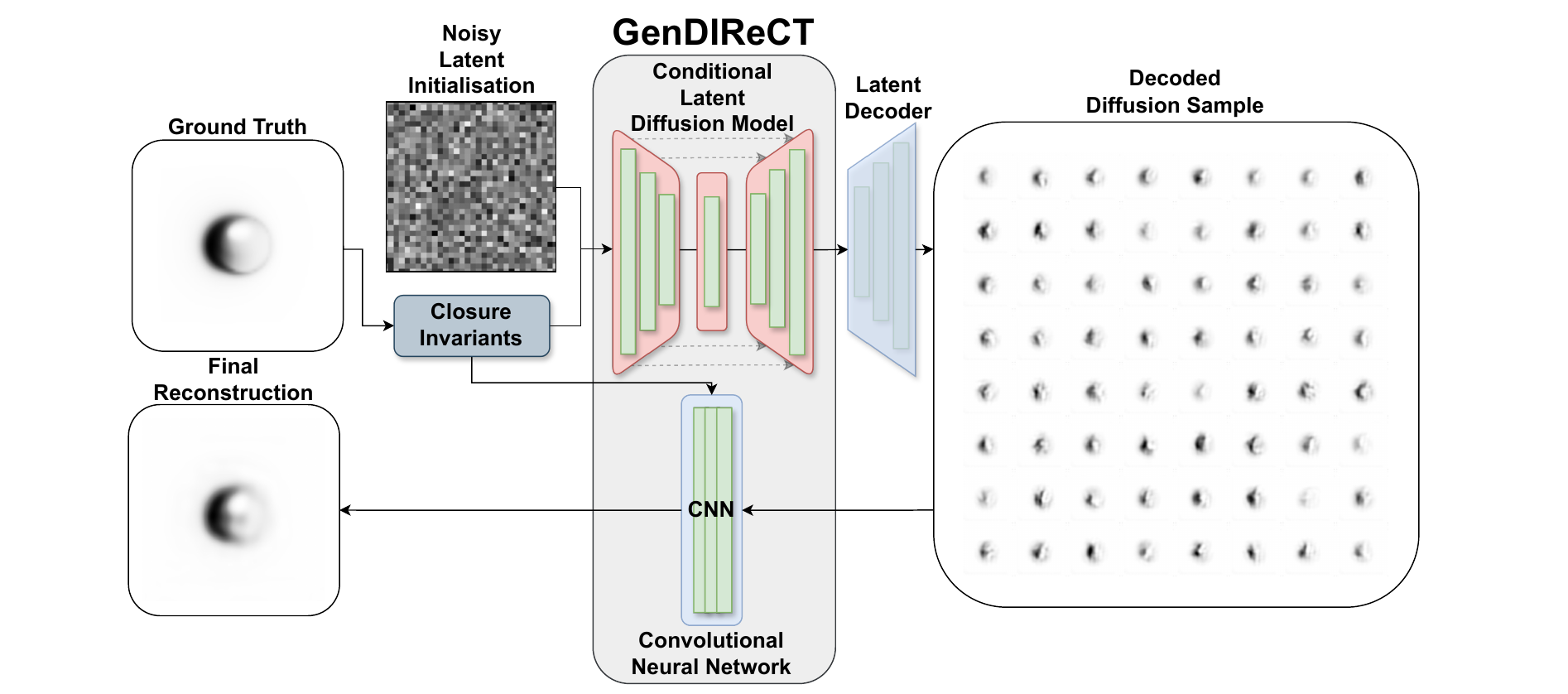}
   \end{tabular}
   \end{center}
   \caption[\texttt{GenDIReCT}] 
   { \label{fig:gendirect} Overview of the \texttt{GenDIReCT} algorithm (reproduced with permission from Lai et al. (2025)\cite{Lai_2025_GenDIReCT}).}
\end{figure}

In order to reduce the reduce the dependence of human-tuned hyperparameters and enable parameter-free imaging, Generative Deep learning Image Reconstruction from Closure Terms (\texttt{GenDIReCT}) \cite{Lai_2025_GenDIReCT} has been developed. \texttt{GenDIReCT} is a generative algorithm trained on non-astronomical images and designed to image directly using only closure invariants. The algorithm employs a latent diffusion model conditioned on closure invariants. \texttt{GenDIReCT} is also unique in that it employs the copolar closure invariants from the general theory of closure invariants \cite{Thyagarajan_2022_CI} which, unlike closure phases and closure amplitudes, presents a unified and homogeneous form for the invariants. The implementation of \texttt{GenDIReCT} is illustrated in Figure~\ref{fig:gendirect}. 

The primary motivation for developing \texttt{GenDIReCT} is to enable an independent analysis of the public EHT observations of M87* and Sgr~A*. Because these datasets have already been extensively analysed by the EHT Collaboration (hereafter, EHTC) and other groups, particular care was taken to minimise the risk of confirmation bias throughout the development process. Rather than tuning the algorithm on the target datasets, we established a hierarchy of validation and performance criteria before attempting any application to EHT data.

Training was performed primarily on non-astronomical images from the CIFAR-10 dataset, supplemented in some cases with simple geometric shapes. This strategy was intentionally adopted to avoid embedding astrophysical expectations into the learned image priors. A key requirement was that the reconstruction framework demonstrate the ability to generalise beyond the image distributions encountered during training and recover structures not explicitly represented in the training set.

To assess robustness, we evaluated consistency across multiple variants of the imaging framework, including models employing different reference stations and different imaging fields of view. Reconstructions were assessed using complementary metrics that probe different aspects of performance, including image-domain similarity measures, adherence to the observed data through closure-invariant residual statistics ($\chi^2$), and the Continuous Ranked Probability Score (CRPS), which is sensitive to multimodality and uncertainty in the solution space. Particular emphasis was placed on ensuring that scientific conclusions remained qualitatively consistent across these independent diagnostics.

A key advantage of the diffusion-based generative framework is its ability to sample the posterior distribution of images conditioned on the closure invariants. Rather than producing a single reconstruction, \texttt{GenDIReCT} generates an ensemble of plausible solutions, enabling direct investigation of reconstruction degeneracies, multimodal structure, and the statistical significance of recovered image features. This provides a principled means of quantifying uncertainty in a highly under-constrained imaging problem.

Only after the algorithm and its hyperparameters satisfied the above validation criteria was the framework applied to the target datasets. No subsequent tuning was performed on the observations that were our scientific targets, thereby preserving the independence of the analysis and reducing the possibility of confirmation bias.

\section{Results}\label{sec:results}

After extensive testing and validation of \texttt{GenDIReCT} \cite{Lai_2025_GenDIReCT}, we applied it to the challenge data set created by the next-generation EHT (\texttt{ngEHT) }collaboration, and the public EHT data on the quasar 3C279, and the radio galaxy Centaurus~A. \texttt{GenDIReCT} also provided the first independent analysis of the public EHT data following the results published by the EHT collaboration on 3C~279 \cite{Kim_2020_3C279} and Centaurus~A (Cen~A) \cite{Janssen_2021_CenA}. This was intended as a staged plan to enhance the confidence in our image reconstruction framework using closure invariants. The results and ongoing work are described below. 

\subsection{Synthetic challenge datasets}\label{sec:synthetic-data}

\begin{figure} [ht]
   \begin{center}
   \begin{tabular}{c} 
   \includegraphics[height=8cm]{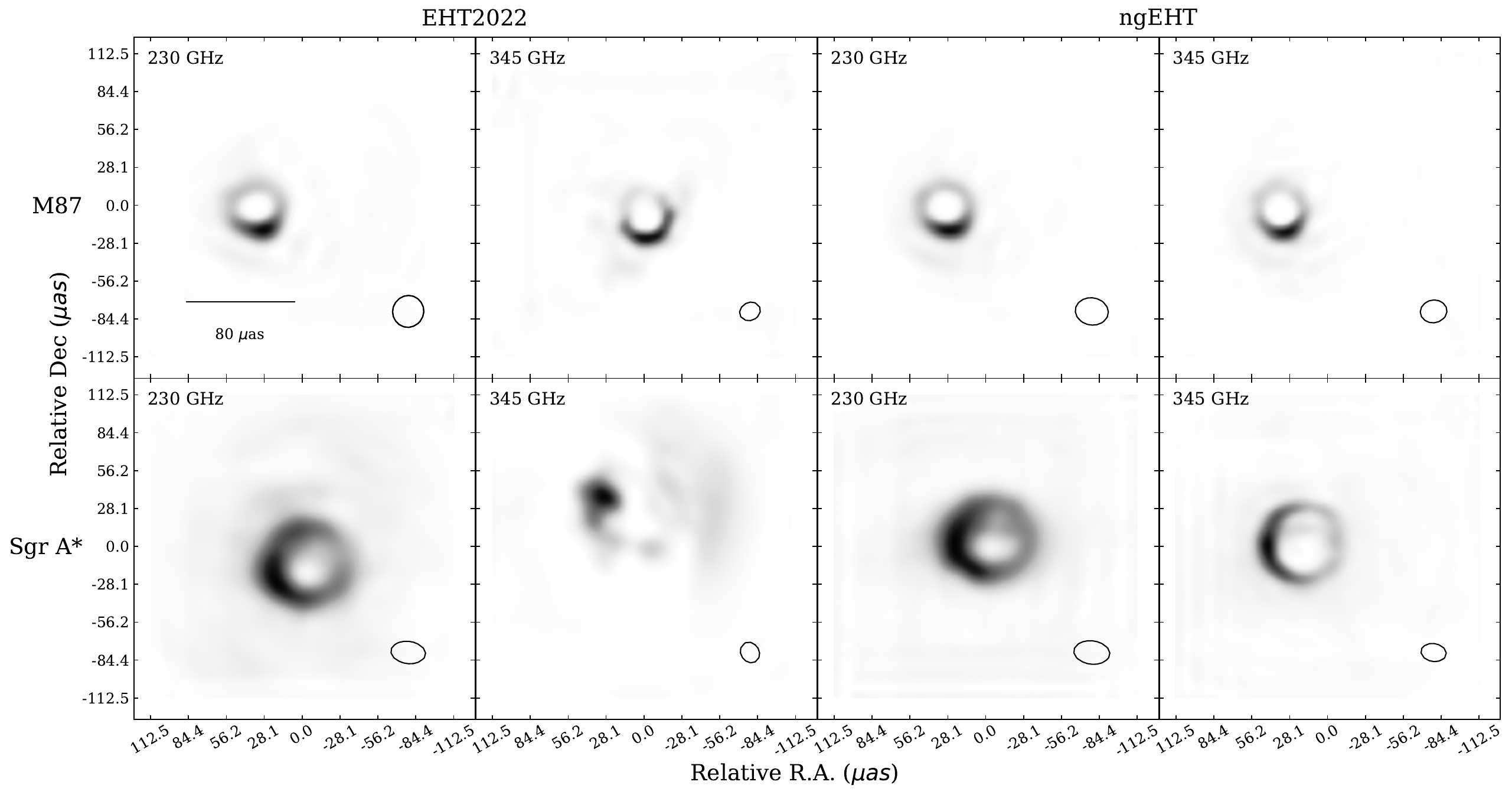}
   \end{tabular}
   \end{center}
   \caption[Challenge dataset] 
   { \label{fig:ngeht-challenge} Results from applying \texttt{GenDIReCT} to the ngEHT challenge datasets. Eight different datasets were used (two bands, two array layouts, and two target source morphologies). This figure is reproduced with permission from Lai et al. (2025) \cite{Lai_2025_GenDIReCT}.}
\end{figure}

In this section, we showcase the results from applying \texttt{GenDIReCT} to ngEHT Analysis Challenges \cite{Roelofs2023_ngeht-challenge}, which are collections of publicly available synthetic VLBI data, designed to advance imaging techniques and algorithms, with a focus on anticipated capability upgrades with the ngEHT array. We adopt the array configuration, observing sequence, and subsequent aperture coverage of the total intensity datasets M87* and Sgr~A* packaged inside the ngEHT Analysis Challenges. The results of our blind application to these datasets are illustrated in Figure~\ref{fig:ngeht-challenge}. It was confirmed that \texttt{GenDIReCT} successfully recovered the structures in all cases except one (Sgr~A* at 345~GHz with the EHT 2022 array configuration). Later, it was confirmed that \texttt{GenDIReCT} achieved comparable or superior results in comparison to other analyses based on the performance metrics set by the team that designed the challenge. We refer the reader to the original paper \cite{Lai_2025_GenDIReCT} for more details. 

\subsection{3C279 data}\label{sec:3C279}

The first application of the \texttt{GenDIReCT} methodology on real EHT observations was the active galactic nuclei 3C~279 \cite{Kim_2020_3C279}, which was observed in 2017 alongside M87. 

\begin{figure} [ht]
   \begin{center}
   \begin{tabular}{c} 
   \includegraphics[height=7cm]{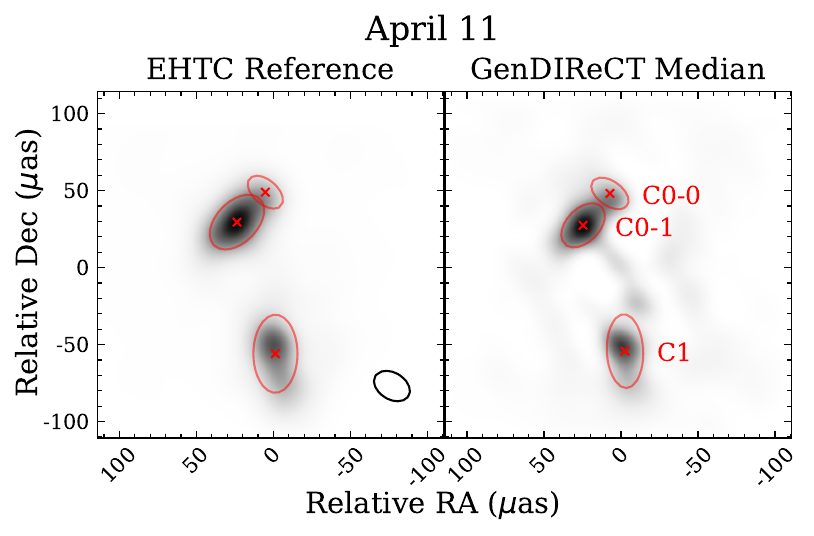}
   \end{tabular}
   \end{center}
   \caption[3C279 EHT data]
   { \label{fig:3C279-image} Results from applying \texttt{GenDIReCT} to the 3C279 EHT data (reproduced with permission from Lai et al. (2026) \cite{Lai_2026_3C279CenA}).}
\end{figure}

The EHT observed 3C~279, interleaved with observations of M87, to independently validate the calibration solution of the M87 image. The data were taken over four nights (5, 6, 10, 11) in April at two 2~GHz bands centered at 227.1~GHz and 229.1~GHz. Additional details of the observation and the EHT collaboration results are described in the EHT 3C~279 paper \cite{Kim_2020_3C279}. \texttt{GenDIReCT} identifies three distinct components in the image (see Figure~\ref{fig:3C279-image}). The image produced by the EHT collaboration is also shown for reference, and it is noted that \texttt{GenDIReCT} finds morphological features that are consistent with the EHTC reference image. 

\begin{figure} [ht]
   \begin{center}
   \begin{tabular}{c} 
   \includegraphics[height=8cm]{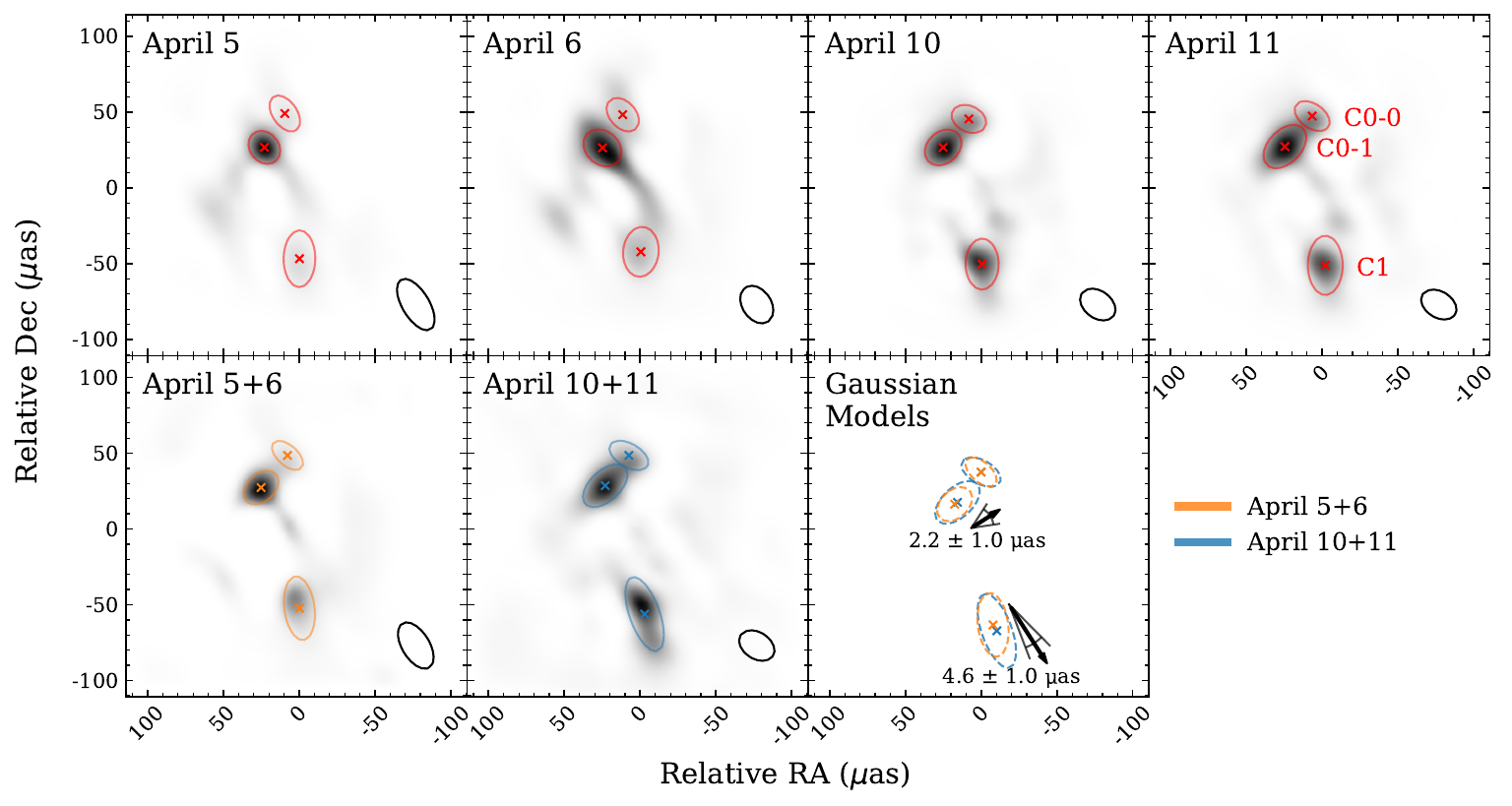}
   \end{tabular}
   \end{center}
   \caption[3C279 proper motion] 
   { \label{fig:3C279-proper-motion} Proper motion of the components in the 3C279 image (reproduced with permission from Lai et al. (2026) \cite{Lai_2026_3C279CenA}).}
\end{figure}

As 3C~279  is known to exhibit apparent superluminal proper motion in one or more of its components \cite{Kim_2020_3C279}, we analysed the data using \texttt{GenDIReCT} separately on each of the days of observations. The proper motion was measured and the apparent superluminal motion was inferred. We found that the southern jet ejecta on sub-parsec scale exhibiting a proper motion of $4.6\pm 1.0\,\mu$as over $\approx 5.39$ days away from the northern components as illustrated in Figure~\ref{fig:3C279-proper-motion}, corresponding to an apparent superluminal velocity of $\simeq (10\pm 2)\,c$. This was also confirmed to be consistent with the EHTC results \cite{Kim_2020_3C279}. 

\subsection{Centaurus A data}\label{sec:CenA}

\begin{figure} [ht]
   \begin{center}
   \begin{tabular}{c} 
   \includegraphics[height=10cm]{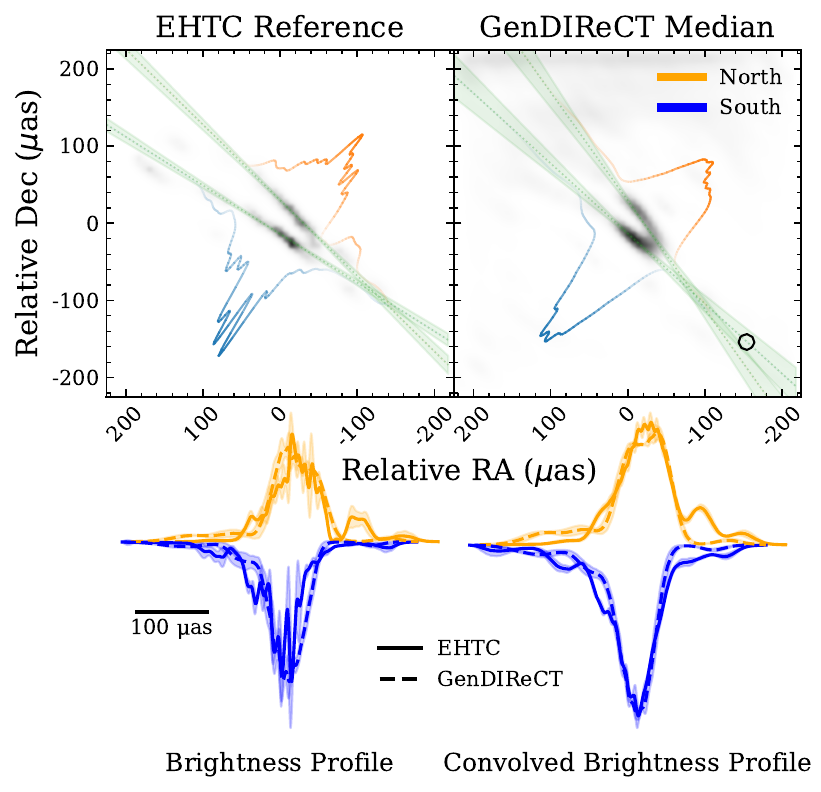}
   \end{tabular}
   \end{center}
   \caption[Centaurus~A data] 
   { \label{fig:CenA-image} Results from applying \texttt{GenDIReCT} to the Centaurus~A EHT data (reproduced with permission from Lai et al. (2026) \cite{Lai_2026_3C279CenA}).}
\end{figure}

Centaurus~A, at a distance of $3.8\pm 0.1$~Mpc \cite{Harris_2010_CenADistance}, is the closest radio galaxy, hosting a supermassive black hole. The EHT observed Centaurus~A for over six hours duration track on 10 April 2017 with both 227.1~GHz and 229.1~GHz bands \cite{Janssen_2021_CenA}. The results from analysing this data using \texttt{GenDIReCT} consistently reconstructed the two bright ridge-lines along the sheath and near the base of the approaching jet as seen in Figure~\ref{fig:CenA-image}. Both the orientation and the opening angle of the ridge lines along the approaching jet are consistent with those reported by the EHTC \cite{Janssen_2021_CenA}. 
 
\subsection{M87* and Sgr~A*}\label{sec:M87-SgrA}

The successes of \texttt{GenDIReCT} strongly motivate its application to our main goals, namely, independently imaging the event horizon scale structures around the M87 and Milky Way black holes. These analyses are underway and will provide the first truly independent analysis of the groundbreaking results that emerged from the EHTC. The independence will not only be in the team and its underlying motivation, but also in the observables and the methodology. 


\section{Summary and future outlook}\label{sec:summary}

The reconstructions presented here and promising ongoing efforts to image the supermassive black hole event horizons in M87 and the Milky Way demonstrate that \texttt{GenDIReCT}, developed and validated through a blind, multi-tier testing framework, provides a robust approach to imaging under the challenging conditions encountered in VLBI. The results further highlight the power of closure invariants as calibration-independent observables that retain rich structural information while remaining resilient to one of the dominant sources of systematic uncertainty in interferometric observations, namely, calibration errors. Together, these advances establish a strong foundation for a new generation of imaging methodologies that place calibration-independent information at the centre of the reconstruction process.

Future developments will extend \texttt{GenDIReCT} beyond the specific characteristics of the EHT. Current efforts are focused on improving the framework's ability to generalise across arbitrary array configurations, observing frequencies, and aperture coverages, thereby reducing or eliminating the need for retraining when applied to new instruments. Such capabilities are particularly important as the astronomical community moves toward increasingly diverse and rapidly evolving interferometric facilities.

The emergence of next-generation VLBI arrays, including expanded EHT networks, the \textit{next-generation Event Horizon Telescope} (ngEHT), and future space-VLBI missions, will dramatically increase both the sensitivity and complexity of event horizon-scale observations. These instruments will probe fainter emission, finer spatial scales, and more rapidly evolving source structures than ever before. Realising their full scientific potential will require imaging frameworks that are not only accurate and computationally scalable, but also robust against calibration uncertainties and capable of rigorously quantifying reconstruction ambiguities.

In this context, approaches such as \texttt{GenDIReCT} offer a valuable complement to traditional imaging methodologies. By combining calibration-independent observables with generative probabilistic inference, they provide an orthogonal pathway for interpreting interferometric data, characterising uncertainties, and assessing the robustness of scientific conclusions. As VLBI continues to push toward the fundamental limits of angular resolution and sensitivity, independent and methodologically distinct imaging frameworks will play an increasingly important role in establishing confidence in some of astronomy's most consequential observations.

 

\bibliography{refs} 

@ARTICLE{Chael_2018_ehtim,
       author = {{Chael}, Andrew A. and {Johnson}, Michael D. and {Bouman}, Katherine L. and {Blackburn}, Lindy L. and {Akiyama}, Kazunori and {Narayan}, Ramesh},
        title = "{Interferometric Imaging Directly with Closure Phases and Closure Amplitudes}",
      journal = {ApJ},
         year = 2018,
        month = apr,
       volume = {857},
       number = {1},
          eid = {23},
        pages = {23},
          doi = {10.3847/1538-4357/aab6a8},
archivePrefix = {arXiv},
       eprint = {1803.07088},
 primaryClass = {astro-ph.IM},
       adsurl = {https://ui.adsabs.harvard.edu/abs/2018ApJ...857...23C}
}

@ARTICLE{Thyagarajan_2022_CPhase,
       author = {{Thyagarajan}, Nithyanandan and {Carilli}, Chris L.},
        title = "{A geometric view of closure phases in interferometry}",
      journal = {PASA},
         year = 2022,
        month = apr,
       volume = {39},
          eid = {e014},
        pages = {e014},
          doi = {10.1017/pasa.2022.6},
archivePrefix = {arXiv},
       eprint = {2012.05254},
 primaryClass = {astro-ph.IM},
       adsurl = {https://ui.adsabs.harvard.edu/abs/2022PASA...39...14T}
}

@ARTICLE{Thyagarajan_2022_CI,
       author = {{Thyagarajan}, Nithyanandan and {Nityananda}, Rajaram and {Samuel}, Joseph},
        title = "{Invariants in copolar interferometry: An Abelian gauge theory}",
      journal = {PRD},
         year = 2022,
        month = feb,
       volume = {105},
       number = {4},
          eid = {043019},
        pages = {043019},
          doi = {10.1103/PhysRevD.105.043019},
archivePrefix = {arXiv},
       eprint = {2108.11399},
 primaryClass = {astro-ph.IM},
       adsurl = {https://ui.adsabs.harvard.edu/abs/2022PhRvD.105d3019T}
}

@BOOK{TMS,
       author = {{Thompson}, A. Richard and {Moran}, James M. and {Swenson}, George W., Jr.},
        title = "{Interferometry and Synthesis in Radio Astronomy, 3rd Edition}",
         year = 2017,
          doi = {10.1007/978-3-319-44431-4},
        publisher = {Astronomy \& Astrophysics Library},
       adsurl = {https://ui.adsabs.harvard.edu/abs/2017isra.book.....T}
}

@ARTICLE{Jennison_1958,
       author = {{Jennison}, R.~C.},
        title = "{A phase sensitive interferometer technique for the measurement of the Fourier transforms of spatial brightness distributions of small angular extent}",
      journal = {MNRAS},
         year = 1958,
        month = jan,
       volume = {118},
        pages = {276},
          doi = {10.1093/mnras/118.3.276},
       adsurl = {https://ui.adsabs.harvard.edu/abs/1958MNRAS.118..276J}
}

@ARTICLE{Twiss_1960,
       author = {{Twiss}, R.~Q. and {Carter}, A.~W.~L. and {Little}, A.~G.},
        title = "{Brightness distribution over some strong radio sources at 1427 Mc/s}",
      journal = {The Observatory},
         year = 1960,
        month = aug,
       volume = {80},
        pages = {153-159},
       adsurl = {https://ui.adsabs.harvard.edu/abs/1960Obs....80..153T}
}

@ARTICLE{Samuel_2022,
       author = {{Samuel}, Joseph and {Nityananda}, Rajaram and {Thyagarajan}, Nithyanandan},
        title = "{Invariants in Polarimetric Interferometry: A Non-Abelian Gauge Theory}",
      journal = {PRL},
         year = 2022,
        month = mar,
       volume = {128},
       number = {9},
          eid = {091101},
        pages = {091101},
          doi = {10.1103/PhysRevLett.128.091101},
archivePrefix = {arXiv},
       eprint = {2108.11400},
 primaryClass = {gr-qc},
       adsurl = {https://ui.adsabs.harvard.edu/abs/2022PhRvL.128i1101S}
}

@ARTICLE{Thyagarajan_2024_Lucas,
       author = {{Thyagarajan}, Nithyanandan and {Hoefs}, Lucas and {Wong}, O. Ivy},
        title = "{Interferometric image reconstruction using closure invariants and machine learning}",
      journal = {RAS Techniques and Instruments},
         year = 2024,
        month = jan,
       volume = {3},
       number = {1},
        pages = {437-452},
          doi = {10.1093/rasti/rzae031},
archivePrefix = {arXiv},
       eprint = {2311.06349},
 primaryClass = {astro-ph.IM},
       adsurl = {https://ui.adsabs.harvard.edu/abs/2024RASTI...3..437T}
}

@ARTICLE{EHT_2022_SgrAImaging,
       author = {{Event Horizon Telescope Collaboration} and {Akiyama}, Kazunori and {Alberdi}, Antxon and {Alef}, Walter and {Algaba}, Juan Carlos and {Anantua}, Richard and {Asada}, Keiichi and {Azulay}, Rebecca and {Bach}, Uwe and {Baczko}, Anne-Kathrin and {Ball}, David and {Balokovi{\'c}}, Mislav and {Barrett}, John and {Baub{\"o}ck}, Michi and {Benson}, Bradford A. and {Bintley}, Dan and {Blackburn}, Lindy and {Blundell}, Raymond and {Bouman}, Katherine L. and {Bower}, Geoffrey C. and {Boyce}, Hope and {Bremer}, Michael and {Brinkerink}, Christiaan D. and {Brissenden}, Roger and {Britzen}, Silke and {Broderick}, Avery E. and {Broguiere}, Dominique and {Bronzwaer}, Thomas and {Bustamante}, Sandra and {Byun}, Do-Young and {Carlstrom}, John E. and {Ceccobello}, Chiara and {Chael}, Andrew and {Chan}, Chi-kwan and {Chatterjee}, Koushik and {Chatterjee}, Shami and {Chen}, Ming-Tang and {Chen}, Yongjun and {Cheng}, Xiaopeng and {Cho}, Ilje and {Christian}, Pierre and {Conroy}, Nicholas S. and {Conway}, John E. and {Cordes}, James M. and {Crawford}, Thomas M. and {Crew}, Geoffrey B. and {Cruz-Osorio}, Alejandro and {Cui}, Yuzhu and {Davelaar}, Jordy and {De Laurentis}, Mariafelicia and {Deane}, Roger and {Dempsey}, Jessica and {Desvignes}, Gregory and {Dexter}, Jason and {Dhruv}, Vedant and {Doeleman}, Sheperd S. and {Dougal}, Sean and {Dzib}, Sergio A. and {Eatough}, Ralph P. and {Emami}, Razieh and {Falcke}, Heino and {Farah}, Joseph and {Fish}, Vincent L. and {Fomalont}, Ed and {Ford}, H. Alyson and {Fraga-Encinas}, Raquel and {Freeman}, William T. and {Friberg}, Per and {Fromm}, Christian M. and {Fuentes}, Antonio and {Galison}, Peter and {Gammie}, Charles F. and {Garc{\'\i}a}, Roberto and {Gentaz}, Olivier and {Georgiev}, Boris and {Goddi}, Ciriaco and {Gold}, Roman and {G{\'o}mez-Ruiz}, Arturo I. and {G{\'o}mez}, Jos{\'e} L. and {Gu}, Minfeng and {Gurwell}, Mark and {Hada}, Kazuhiro and {Haggard}, Daryl and {Haworth}, Kari and {Hecht}, Michael H. and {Hesper}, Ronald and {Heumann}, Dirk and {Ho}, Luis C. and {Ho}, Paul and {Honma}, Mareki and {Huang}, Chih-Wei L. and {Huang}, Lei and {Hughes}, David H. and {Ikeda}, Shiro and {Impellizzeri}, C.~M. Violette and {Inoue}, Makoto and {Issaoun}, Sara and {James}, David J. and {Jannuzi}, Buell T. and {Janssen}, Michael and {Jeter}, Britton and {Jiang}, Wu and {Jim{\'e}nez-Rosales}, Alejandra and {Johnson}, Michael D. and {Jorstad}, Svetlana and {Joshi}, Abhishek V. and {Jung}, Taehyun and {Karami}, Mansour and {Karuppusamy}, Ramesh and {Kawashima}, Tomohisa and {Keating}, Garrett K. and {Kettenis}, Mark and {Kim}, Dong-Jin and {Kim}, Jae-Young and {Kim}, Jongsoo and {Kim}, Junhan and {Kino}, Motoki and {Koay}, Jun Yi and {Kocherlakota}, Prashant and {Kofuji}, Yutaro and {Koch}, Patrick M. and {Koyama}, Shoko and {Kramer}, Carsten and {Kramer}, Michael and {Krichbaum}, Thomas P. and {Kuo}, Cheng-Yu and {La Bella}, Noemi and {Lauer}, Tod R. and {Lee}, Daeyoung and {Lee}, Sang-Sung and {Leung}, Po Kin and {Levis}, Aviad and {Li}, Zhiyuan and {Lico}, Rocco and {Lindahl}, Greg and {Lindqvist}, Michael and {Lisakov}, Mikhail and {Liu}, Jun and {Liu}, Kuo and {Liuzzo}, Elisabetta and {Lo}, Wen-Ping and {Lobanov}, Andrei P. and {Loinard}, Laurent and {Lonsdale}, Colin J. and {Lu}, Ru-Sen and {Mao}, Jirong and {Marchili}, Nicola and {Markoff}, Sera and {Marrone}, Daniel P. and {Marscher}, Alan P. and {Mart{\'\i}-Vidal}, Iv{\'a}n and {Matsushita}, Satoki and {Matthews}, Lynn D. and {Medeiros}, Lia and {Menten}, Karl M. and {Michalik}, Daniel and {Mizuno}, Izumi and {Mizuno}, Yosuke and {Moran}, James M. and {Moriyama}, Kotaro and {Moscibrodzka}, Monika and {M{\"u}ller}, Cornelia and {Mus}, Alejandro and {Musoke}, Gibwa and {Myserlis}, Ioannis and {Nadolski}, Andrew and {Nagai}, Hiroshi and {Nagar}, Neil M. and {Nakamura}, Masanori and {Narayan}, Ramesh and {Narayanan}, Gopal and {Natarajan}, Iniyan and {Nathanail}, Antonios and {Fuentes}, Santiago Navarro and {Neilsen}, Joey and {Neri}, Roberto and {Ni}, Chunchong and {Noutsos}, Aristeidis and {Nowak}, Michael A. and {Oh}, Junghwan and {Okino}, Hiroki and {Olivares}, H{\'e}ctor and {Ortiz-Le{\'o}n}, Gisela N. and {Oyama}, Tomoaki and {{\"O}zel}, Feryal and {Palumbo}, Daniel C.~M. and {Paraschos}, Georgios Filippos and {Park}, Jongho and {Parsons}, Harriet and {Patel}, Nimesh and {Pen}, Ue-Li and {Pesce}, Dominic W. and {Pi{\'e}tu}, Vincent and {Plambeck}, Richard and {PopStefanija}, Aleksandar and {Porth}, Oliver and {P{\"o}tzl}, Felix M. and {Prather}, Ben and {Preciado-L{\'o}pez}, Jorge A. and {Psaltis}, Dimitrios and {Pu}, Hung-Yi and {Ramakrishnan}, Venkatessh and {Rao}, Ramprasad and {Rawlings}, Mark G. and {Raymond}, Alexander W. and {Rezzolla}, Luciano and {Ricarte}, Angelo and {Ripperda}, Bart and {Roelofs}, Freek and {Rogers}, Alan and {Ros}, Eduardo and {Romero-Ca{\~n}izales}, Cristina and {Roshanineshat}, Arash and {Rottmann}, Helge and {Roy}, Alan L. and {Ruiz}, Ignacio and {Ruszczyk}, Chet and {Rygl}, Kazi L.~J. and {S{\'a}nchez}, Salvador and {S{\'a}nchez-Arg{\"u}elles}, David and {S{\'a}nchez-Portal}, Miguel and {Sasada}, Mahito and {Satapathy}, Kaushik and {Savolainen}, Tuomas and {Schloerb}, F. Peter and {Schonfeld}, Jonathan and {Schuster}, Karl-Friedrich and {Shao}, Lijing and {Shen}, Zhiqiang and {Small}, Des and {Sohn}, Bong Won and {SooHoo}, Jason and {Souccar}, Kamal and {Sun}, He and {Tazaki}, Fumie and {Tetarenko}, Alexandra J. and {Tiede}, Paul and {Tilanus}, Remo P.~J. and {Titus}, Michael and {Torne}, Pablo and {Traianou}, Efthalia and {Trent}, Tyler and {Trippe}, Sascha and {Turk}, Matthew and {van Bemmel}, Ilse and {van Langevelde}, Huib Jan and {van Rossum}, Daniel R. and {Vos}, Jesse and {Wagner}, Jan and {Ward-Thompson}, Derek and {Wardle}, John and {Weintroub}, Jonathan and {Wex}, Norbert and {Wharton}, Robert and {Wielgus}, Maciek and {Wiik}, Kaj and {Witzel}, Gunther and {Wondrak}, Michael F. and {Wong}, George N. and {Wu}, Qingwen and {Yamaguchi}, Paul and {Yoon}, Doosoo and {Young}, Andr{\'e} and {Young}, Ken and {Younsi}, Ziri and {Yuan}, Feng and {Yuan}, Ye-Fei and {Zensus}, J. Anton and {Zhang}, Shuo and {Zhao}, Guang-Yao and {Zhao}, Shan-Shan},
        title = "{First Sagittarius A* Event Horizon Telescope Results. III. Imaging of the Galactic Center Supermassive Black Hole}",
      journal = {ApJL},
         year = 2022,
        month = may,
       volume = {930},
       number = {2},
          eid = {L14},
        pages = {L14},
          doi = {10.3847/2041-8213/ac6429},
       adsurl = {https://ui.adsabs.harvard.edu/abs/2022ApJ...930L..14E}
}

@INPROCEEDINGS{Doeleman_2009,
       author = {{Doeleman}, Sheperd and {Agol}, Eric and {Backer}, Don and {Baganoff}, Fred and {Bower}, Geoffrey C. and {Broderick}, Avery and {Fabian}, Andrew and {Fish}, Vincent and {Gammie}, Charles and {Ho}, Paul and {Honman}, Mareki and {Krichbaum}, Thomas and {Loeb}, Avi and {Marrone}, Dan and {Reid}, Mark and {Rogers}, Alan and {Shapiro}, Irwin and {Strittmatter}, Peter and {Tilanus}, Remo and {Weintroub}, Jonathan and {Whitney}, Alan and {Wright}, Melvyn and {Ziurys}, Lucy},
        title = "{Imaging an Event Horizon: submm-VLBI of a Super Massive Black Hole}",
    booktitle = {astro2010: The Astronomy and Astrophysics Decadal Survey},
         year = 2009,
       volume = {2010},
        month = jan,
        pages = {68},
          doi = {10.48550/arXiv.0906.3899},
archivePrefix = {arXiv},
       eprint = {0906.3899},
 primaryClass = {astro-ph.CO},
       adsurl = {https://ui.adsabs.harvard.edu/abs/2009astro2010S..68D}
}

@ARTICLE{Broderick_2022,
       author = {{Broderick}, Avery E. and {Pesce}, Dominic W. and {Gold}, Roman and {Tiede}, Paul and {Pu}, Hung-Yi and {Anantua}, Richard and {Britzen}, Silke and {Ceccobello}, Chiara and {Chatterjee}, Koushik and {Chen}, Yongjun and {Conroy}, Nicholas S. and {Crew}, Geoffrey B. and {Cruz-Osorio}, Alejandro and {Cui}, Yuzhu and {Doeleman}, Sheperd S. and {Emami}, Razieh and {Farah}, Joseph and {Fromm}, Christian M. and {Galison}, Peter and {Georgiev}, Boris and {Ho}, Luis C. and {James}, David J. and {Jeter}, Britton and {Jimenez-Rosales}, Alejandra and {Koay}, Jun Yi and {Kramer}, Carsten and {Krichbaum}, Thomas P. and {Lee}, Sang-Sung and {Lindqvist}, Michael and {Mart{\'\i}-Vidal}, Iv{\'a}n and {Menten}, Karl M. and {Mizuno}, Yosuke and {Moran}, James M. and {Moscibrodzka}, Monika and {Nathanail}, Antonios and {Neilsen}, Joey and {Ni}, Chunchong and {Park}, Jongho and {Pi{\'e}tu}, Vincent and {Rezzolla}, Luciano and {Ricarte}, Angelo and {Ripperda}, Bart and {Shao}, Lijing and {Tazaki}, Fumie and {Toma}, Kenji and {Torne}, Pablo and {Weintroub}, Jonathan and {Wielgus}, Maciek and {Yuan}, Feng and {Zhao}, Shan-Shan and {Zhang}, Shuo},
        title = "{The Photon Ring in M87*}",
      journal = {ApJ},
         year = 2022,
        month = aug,
       volume = {935},
       number = {1},
          eid = {61},
        pages = {61},
          doi = {10.3847/1538-4357/ac7c1d},
archivePrefix = {arXiv},
       eprint = {2208.09004},
 primaryClass = {astro-ph.HE},
       adsurl = {https://ui.adsabs.harvard.edu/abs/2022ApJ...935...61B}
}

@ARTICLE{Janssen_2021_CenA,
       author = {{Janssen}, Michael and {Falcke}, Heino and {Kadler}, Matthias and {Ros}, Eduardo and {Wielgus}, Maciek and {Akiyama}, Kazunori and {Balokovi{\'c}}, Mislav and {Blackburn}, Lindy and {Bouman}, Katherine L. and {Chael}, Andrew and {Chan}, Chi-kwan and {Chatterjee}, Koushik and {Davelaar}, Jordy and {Edwards}, Philip G. and {Fromm}, Christian M. and {G{\'o}mez}, Jos{\'e} L. and {Goddi}, Ciriaco and {Issaoun}, Sara and {Johnson}, Michael D. and {Kim}, Junhan and {Koay}, Jun Yi and {Krichbaum}, Thomas P. and {Liu}, Jun and {Liuzzo}, Elisabetta and {Markoff}, Sera and {Markowitz}, Alex and {Marrone}, Daniel P. and {Mizuno}, Yosuke and {M{\"u}ller}, Cornelia and {Ni}, Chunchong and {Pesce}, Dominic W. and {Ramakrishnan}, Venkatessh and {Roelofs}, Freek and {Rygl}, Kazi L.~J. and {van Bemmel}, Ilse and {Event Horizon Telescope Collaboration} and {Alberdi}, Antxon and {Alef}, Walter and {Algaba}, Juan Carlos and {Anantua}, Richard and {Asada}, Keiichi and {Azulay}, Rebecca and {Baczko}, Anne-Kathrin and {Ball}, David and {Ball}, David and {Barrett}, John and {Benson}, Bradford A. and {Bintley}, Dan and {Bintley}, Dan and {Blundell}, Raymond and {Boland}, Wilfred and {Boland}, Wilfred and {Bower}, Geoffrey C. and {Boyce}, Hope and {Bremer}, Michael and {Brinkerink}, Christiaan D. and {Brissenden}, Roger and {Britzen}, Silke and {Broderick}, Avery E. and {Broguiere}, Dominique and {Bronzwaer}, Thomas and {Byun}, Do-Young and {Carlstrom}, John E. and {Chatterjee}, Shami and {Chen}, Ming-Tang and {Chen}, Yongjun and {Chesler}, Paul M. and {Cho}, Ilje and {Christian}, Pierre and {Conway}, John E. and {Cordes}, James M. and {Crawford}, Thomas M. and {Crew}, Geoffrey B. and {Cruz-Osorio}, Alejandro and {Cui}, Yuzhu and {Cui}, Yuzhu and {De Laurentis}, Mariafelicia and {Deane}, Roger and {Dempsey}, Jessica and {Desvignes}, Gregory and {Dexter}, Jason and {Doeleman}, Sheperd S. and {Eatough}, Ralph P. and {Farah}, Joseph and {Farah}, Joseph and {Fish}, Vincent L. and {Fomalont}, Ed and {Ford}, H. Alyson and {Fraga-Encinas}, Raquel and {Friberg}, Per and {Friberg}, Per and {Fuentes}, Antonio and {Galison}, Peter and {Gammie}, Charles F. and {Garc{\'\i}a}, Roberto and {Gelles}, Zachary and {Gentaz}, Olivier and {Georgiev}, Boris and {Georgiev}, Boris and {Gold}, Roman and {Gold}, Roman and {G{\'o}mez-Ruiz}, Arturo I. and {Gu}, Minfeng and {Gurwell}, Mark and {Hada}, Kazuhiro and {Haggard}, Daryl and {Hecht}, Michael H. and {Hesper}, Ronald and {Himwich}, Elizabeth and {Ho}, Luis C. and {Ho}, Paul and {Honma}, Mareki and {Huang}, Chih-Wei L. and {Huang}, Lei and {Hughes}, David H. and {Ikeda}, Shiro and {Inoue}, Makoto and {Inoue}, Makoto and {James}, David J. and {Jannuzi}, Buell T. and {Jeter}, Britton and {Jiang}, Wu and {Jimenez-Rosales}, Alejandra and {Jorstad}, Svetlana and {Jung}, Taehyun and {Karami}, Mansour and {Karuppusamy}, Ramesh and {Kawashima}, Tomohisa and {Keating}, Garrett K. and {Kettenis}, Mark and {Kim}, Dong-Jin and {Kim}, Jae-Young and {Kim}, Jongsoo and {Kino}, Motoki and {Kofuji}, Yutaro and {Koyama}, Shoko and {Kramer}, Michael and {Kramer}, Carsten and {Kuo}, Cheng-Yu and {Lauer}, Tod R. and {Lee}, Sang-Sung and {Levis}, Aviad and {Li}, Yan-Rong and {Li}, Zhiyuan and {Lindqvist}, Michael and {Lico}, Rocco and {Lindahl}, Greg and {Liu}, Kuo and {Lo}, Wen-Ping and {Lobanov}, Andrei P. and {Loinard}, Laurent and {Lonsdale}, Colin and {Lu}, Ru-Sen and {MacDonald}, Nicholas R. and {Mao}, Jirong and {Marchili}, Nicola and {Marscher}, Alan P. and {Mart{\'\i}-Vidal}, Iv{\'a}n and {Matsushita}, Satoki and {Matthews}, Lynn D. and {Medeiros}, Lia and {Menten}, Karl M. and {Mizuno}, Izumi and {Moran}, James M. and {Moriyama}, Kotaro and {Moscibrodzka}, Monika and {Moscibrodzka}, Monika and {Musoke}, Gibwa and {Mej{\'\i}as}, Alejandro Mus and {Nagai}, Hiroshi and {Nagar}, Neil M. and {Nakamura}, Masanori and {Narayan}, Ramesh and {Narayanan}, Gopal and {Natarajan}, Iniyan and {Nathanail}, Antonios and {Neilsen}, Joey and {Neri}, Roberto and {Noutsos}, Aristeidis and {Nowak}, Michael A. and {Okino}, Hiroki and {Olivares}, H{\'e}ctor and {Ortiz-Le{\'o}n}, Gisela N. and {Oyama}, Tomoaki and {{\"O}zel}, Feryal and {Palumbo}, Daniel C.~M. and {Park}, Jongho and {Patel}, Nimesh and {Pen}, Ue-Li and {Pi{\'e}tu}, Vincent and {Plambeck}, Richard and {PopStefanija}, Aleksandar and {Porth}, Oliver and {P{\"o}tzl}, Felix M. and {Prather}, Ben and {Preciado-L{\'o}pez}, Jorge A. and {Psaltis}, Dimitrios and {Pu}, Hung-Yi and {Pu}, Hung-Yi and {Rao}, Ramprasad and {Rawlings}, Mark G. and {Raymond}, Alexander W. and {Rezzolla}, Luciano and {Ricarte}, Angelo and {Ripperda}, Bart and {Rogers}, Alan and {Rose}, Mel and {Roshanineshat}, Arash and {Rottmann}, Helge and {Roy}, Alan L. and {Ruszczyk}, Chet and {S{\'a}nchez}, Salvador and {S{\'a}nchez-Arguelles}, David and {Sasada}, Mahito and {Savolainen}, Tuomas and {Schloerb}, F. Peter and {Schuster}, Karl-Friedrich and {Shao}, Lijing and {Shen}, Zhiqiang and {Small}, Des and {Sohn}, Bong Won and {SooHoo}, Jason and {Sun}, He and {Tazaki}, Fumie and {Tetarenko}, Alexandra J. and {Tiede}, Paul and {Tilanus}, Remo P.~J. and {Titus}, Michael and {Torne}, Pablo and {Trent}, Tyler and {Traianou}, Efthalia and {Trippe}, Sascha and {van Bemmel}, Ilse and {van Langevelde}, Huib Jan and {van Rossum}, Daniel R. and {Wagner}, Jan and {Ward-Thompson}, Derek and {Wardle}, John and {Weintroub}, Jonathan and {Wex}, Norbert and {Wharton}, Robert and {Wharton}, Robert and {Wong}, George N. and {Wu}, Qingwen and {Yoon}, Doosoo and {Young}, Andr{\'e} and {Young}, Ken and {Younsi}, Ziri and {Yuan}, Feng and {Yuan}, Ye-Fei and {Zensus}, J. Anton and {Zhao}, Guang-Yao and {Zhao}, Shan-Shan},
        title = "{Event Horizon Telescope observations of the jet launching and collimation in Centaurus A}",
      journal = {Nature Astronomy},
         year = 2021,
        month = jul,
       volume = {5},
        pages = {1017-1028},
          doi = {10.1038/s41550-021-01417-w},
archivePrefix = {arXiv},
       eprint = {2111.03356},
 primaryClass = {astro-ph.GA},
       adsurl = {https://ui.adsabs.harvard.edu/abs/2021NatAs...5.1017J}
}

@ARTICLE{Lai_2025_GenDIReCT,
       author = {{Lai}, Samuel and {Thyagarajan}, Nithyanandan and {Wong}, O. Ivy and {Diakogiannis}, Foivos},
        title = "{Very-long baseline interferometry imaging with closure invariants using conditional image diffusion}",
      journal = {PASA},
         year = 2025,
        month = nov,
       volume = {42},
          eid = {e148},
        pages = {e148},
          doi = {10.1017/pasa.2025.10110},
archivePrefix = {arXiv},
       eprint = {2510.12093},
 primaryClass = {astro-ph.IM},
       adsurl = {https://ui.adsabs.harvard.edu/abs/2025PASA...42..148L}
}

@ARTICLE{Roelofs2023_ngeht-challenge,
       author = {{Roelofs}, Freek and {Blackburn}, Lindy and {Lindahl}, Greg and {Doeleman}, Sheperd S. and {Johnson}, Michael D. and {Arras}, Philipp and {Chatterjee}, Koushik and {Emami}, Razieh and {Fromm}, Christian and {Fuentes}, Antonio and {Knollm{\"u}ller}, Jakob and {Kosogorov}, Nikita and {M{\"u}ller}, Hendrik and {Patel}, Nimesh and {Raymond}, Alexander and {Tiede}, Paul and {Traianou}, Efthalia and {Vega}, Justin},
        title = "{The ngEHT Analysis Challenges}",
      journal = {Galaxies},
         year = 2023,
        month = jan,
       volume = {11},
       number = {1},
          eid = {12},
        pages = {12},
          doi = {10.3390/galaxies11010012},
archivePrefix = {arXiv},
       eprint = {2212.11355},
 primaryClass = {astro-ph.IM},
       adsurl = {https://ui.adsabs.harvard.edu/abs/2023Galax..11...12R}
}
\bibliographystyle{spiebib} 

\end{document}